\documentclass{article}
\usepackage[utf8]{inputenc} 
\usepackage{graphicx}
\usepackage[legalpaper, margin=0.9in]{geometry}

\usepackage{comment}
\usepackage{times}
\usepackage{graphicx}
 \usepackage{url}
 \usepackage{xcolor}
 \usepackage{tcolorbox}
\usepackage{color, colortbl}
\usepackage{graphicx,amsmath,amssymb}
\usepackage{soul}
\DeclareRobustCommand{\bdiamond}{%
  \mathbin{\text{\scalebox{.75}{\rotatebox[origin=c]{45}{$\Box$}}}}%
}

\usepackage{todonotes}
\usepackage{amsmath}
\usepackage{multicol}
\usepackage{comment}
\usepackage{authblk}
\usepackage[bottom]{footmisc}

\title{A collaborative path to scientific discovery: Distribution of labor, productivity and innovation in collaborative science}
\date{} 
\author[1]{Floriana Gargiulo \footnote{Correspondence: floriana.gargiulo@cnrs.fr}}
\author[2]{Maria Castaldo}
\author[3]{Tommaso Venturini}
\author[2]{Paolo Frasca}

\affil[1]{CNRS, Université Paris-Sorbonne - Paris IV, GEMASS,
  59 rue Pouchet,
  F-75017,
  Paris,
  France}
\affil[2]{Univ.\ Grenoble Alpes, CNRS, Inria, Grenoble INP, GIPSA-lab,
  11 rue des Mathématiques,
  F-38000,
  Grenoble,
  France}

\affil[3]{CNRS, CIS-lab,
  59 rue Pouchet,
  F-75017,
  Paris,
  France}

\begin{document}

\maketitle




{ \begin{abstract}
In this work we dig into the process of scientific discovery by looking at a yet unexploited source of information: Polymath projects. Polymath projects are an original attempt to collectively solve mathematical problems in an online collaborative environment.
To investigate the Polymath experiment, we analyze all the posts related to the projects that arrived to a peer reviewed publication with a particular attention to the organization of labor and the innovations originating from the author contributions. We observe that a significant presence of sporadic contributor boosts the productivity of the most active users and that  productivity, in terms of number of posts, grows super-linearly with the number of contributors. When it comes to innovation in large scale collaborations, there is no exact rule determining, a priori, who the main innovators will be. Sometimes, serendipitous interactions by sporadic contributors can have a large impact on the discovery process and a single post by an occasional participant can steer the work into a new direction.
\end{abstract}}

\section*{Introduction}
\label{intro}
The history of massive online collaboration began in 1991, when Linus Torvald introduced Linux, the first example of a fast growing collaborative project. After Linux, the development of advanced techniques of versioning provided a working architecture for large scale collaborations and gave way to the development of Wikipedia, the largest human collaborative project, and of the social coding platform GitHub. These platforms have also shown an enormous research potential for studying collaboration patterns and more generally for analyzing  human behavior. A large number of scholars from multiple disciplines analyzed patterns of users' interactions in large--scale collaborative systems like Linux \cite{maillartLinux}, Wikipedia \cite{voss,yasseri1,ciampaglia,yasseri2} and GitHub \cite{thung,guzman,sornette}. \\
These collaborative experiences express a deep transformation of the organisation of the production of code and knowledge: a shift from the "waterfall model", where collaboration is hierarchically organized with a static division of labor where each actor has a precise role, to the so called collective intelligence, where the actors have no assigned roles and can randomly participate to the tasks they prefer \cite{malone,benkler}. 
The expression “wisdom of the crowds” \cite{galton} describes the, sometime surprising, creative power generated by this type of self-organized interactions \cite{becker,kittur}.
This collective wisdom emerges thanks to a sort of designed serendipity promoted by the platforms: the diversity of the participants, and their micro-expertise, can timely intervene to unlock the creative process and generate cascades of major innovations. \\
 Success stories such as the one of Linux or Wikipedia question the idea of inventions as the brainchild of individual minds and instead suggest that cultural achievements can instead be the results of a multiplicity of invisible contributions without which no stroke of genius would be impossible. Such a conceptual framing has vast implications when applied to scientific activities: is science the craft of many or of a few? Can research  be carried out in a large-scale open collaborative environment? Can science be based on collaboration instead of competition? These questions are discussed in the book “Reinventing discovery” by Michael Nielsen \cite{nielsen}, which presents several instances of collective "problem solving" challenges. Among the cases described by Nielsen, one -- the Polymath project -- caught our attention as it can be studied not only ethnographically, but also computationally since all the contributions that made it up took place in writing and are available as digital records.\\
The first Polymath project was proposed in 2009 by the mathematician Tim Gowers with a post on his blog where he invited all the mathematicians to contribute to the solution of a problem in number theory, using a dedicated discussion thread on his blog. 
After this other 15 polymath projects followed and 6 of these arrived to a final peer reviewed publication signed with the collective name “polymath collaboration”. 
The Polymath ecosystem not only allows to study collaborative science but it also constitutes an unprecedented playground for an in-depth study of the discovery processes.
A preliminary descriptive analysis of the Polymath 1 project has been presented in \cite{barany,cranshaw}  {--} to our knowledge, the only scientific papers that have investigated this collaborative experience, along with the reflection on the Polymath Projects done by Tim Gowers himself  \cite{gowers_massively_2009}.

In our work we present a rich statistical analysis of the Polymath ecosystem, addressing in particular the participants’ activity and the contents they produced. Our findings are based on the analysis of those projects that arrived to a final peer-reviewed publication (projects 1,4,5,8,15). In section 1 we present the data and methods used in our analysis and in section 2 we present our findings. Sections \ref{laborOrg} and \ref{collInt} analyse the internal structure of collaboration and its role in productivity patterns. 
We identify a clear hierarchy in participation patterns with an hyper-active elite responsible for 80\% of the work. We show that the collaborative architecture had a major role in fostering the individual production: we observe super--production patterns by which the presence of occasional participants contributes to increase the productivity of the elite. \\
After analyzing the open-science labor organization, we focus on the analysis of the mechanisms of scientific discovery. A mathematical discovery consists in the proof of a set of hypothesis realized merging together a set of pre-existing theorems, conjectures, axioms, etc. It enters therefore in a larger framework of innovative processes where the introduction of novel concepts can be fundamental to finalize certain proofs. Innovation processes can be described by the notion of {\em adjacent possible expansion}, introduced by Stuart Kauffman \cite{kauffman}, who refers to the continuous expansion, or restructuring, of the possible knowledge space, triggered by the coming into play of novel concepts. It has been shown that this kind of processes leaves clear markers in the statistical properties of the produced knowledge, expressed by two important laws observed for the first time in linguistic: Zipf's law and Heaps' law \cite{loretoEntropy}. In section \ref{statProp} we show that mathematical discovery mechanisms exhibit the markers of the adjacent possible expansion processes, akin to literature production and music innovation.
Finally, in section \ref{innovPatterns} we investigate trigger factors for innovation and we show that there is no exact rule determining, a priori, who the main innovators will be. In fact, not only major contributors but also peripheral users can, in some cases, steer the collective work into new directions. 

\section{Data and Methods}
\label{sec:1} 
We collected all the posts of Polymath projects 1, 4, 5, 8 and 15 starting from the links collected on the Wiki page of the Polymath project \cite{wiki}. The corpus for each project is constituted by a collection of posts, identified by their date, their author, the text and the parent post if they were written as a reply to another post. Posts were published mainly on three blogs: Timothy Gowers's \cite{gowersblog}, Terence Tao's \cite{taoblog} and the Polymath blog \cite{polymathblog}. Each of these blogs introduces different constraints in the way authors interact. In particular, in Timothy Gowers's blog, comments can only be posted on main threads, hence the depth of the discussion is limited and authors cannot react to comments of main threads. Terence Tao's blog, on the other hand, allows comments up to a depth equal to 4. The Polymath blog does not seem to limit nested comments of any level and shows comments up to a depth equal to 10.

\subsection{Demography of the projects}
\begin{figure*}
\includegraphics[width=\textwidth]{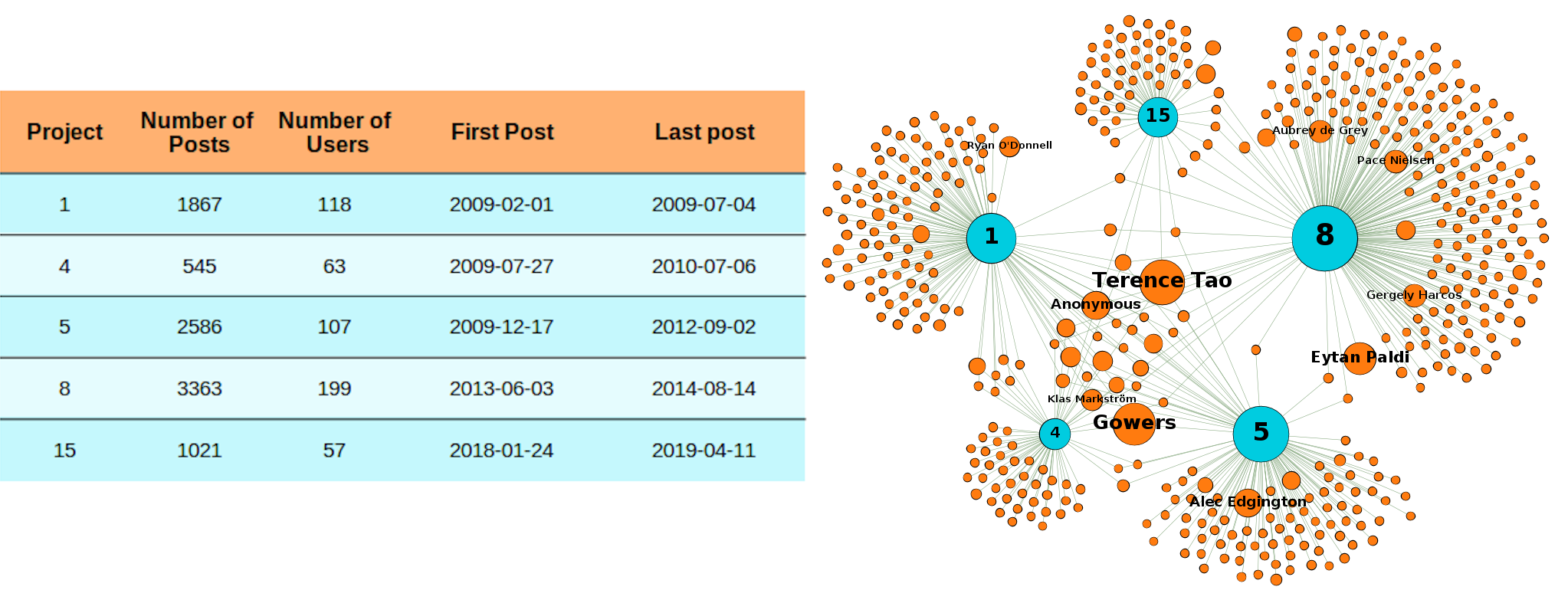}
\caption{The table on the left shows the number of posts, the number of contributors and the periods of development for each project. The network on the right represents the participation of each author to the projects.}
\label{demography}       
\end{figure*}

The five projects we analysed contain between 545 and 3363 posts and the number of contributors varies between 57 and 199. The details of each project are reported in the table in Fig~\ref{demography}. The network in Fig~\ref{demography} represents the bipartite network of contributors  and projects: in the graph an edge represents the participation of a contributor in a project. The size of the contributors' nodes represents the number of posts they published. The chart reveals the presence of a small core of very active authors who participated to almost all projects and a periphery of minor contributors working on one project only.  
\subsection{Contents' identification}
Since we are interested in reconstructing the collaborative processes that solved the proposed problems, we need to identify the mathematical objects  {used} by authors in their posts. Natural language processing techniques performed poorly on this task and tended to identify non-mathematical terms such as characteristics of the personal linguistic patterns of each contributor. Hence, we built a mathematical vocabulary through a two steps process. First, we collected the titles of all the Wikipedia pages labeled as mathematics \cite{wikimath}. Second we added to the list all the expressions "Theorem of *", "*'s conjecture", etc. extracted from the corpora. 
\\The dictionary we obtained contains 25035 mathematical concepts. Table \ref{tab:1} shows the number of independents mathematical concepts retrieved in the projects: when an expression like "theorem of *" appears, the sub-string "theorem" is not considered as an independent concept.  Using the mathematical dictionary, the content of each post is identified as the set of mathematical concepts it contains: $K_i=\{kw_1,...,kw_m\}$. A post will be generally characterized by its time ($t$), its author ($\alpha$) and its content ($K$): $\Pi_i=(t_i,\alpha_i,K_i)$.

\begin{table}[h!]
\centering
\label{tab:1}       
\begin{tabular}{ll}
\hline\noalign{\smallskip}
Project & Concepts \\
\noalign{\smallskip}\hline\noalign{\smallskip}
1 & 1040  \\
4 & 768 \\
5 & 1089 \\
8 & 1116 \\
15 & 661 \\
\noalign{\smallskip}\hline
\end{tabular}
\caption{Number of independent mathematical concepts in the corpora}
\end{table}

\subsection{Topic extraction}
We then aggregate different mathematical concepts into topics. This aggregation allows us to investigate the collaborations between authors and the structure of labor. To define topics, for each Polymath project, we build a co-occurrences network, where nodes represent different mathematical concepts and they are linked to each other if there exists at least a post in which they were discussed together. The network is weighted and the weight associated to each link is the number of co-occurrences in different posts. As the obtained network is highly connected and extremely complex, we filter the edges to  {identify} relevant structures. In order to do so, we compute the Planar Maximally Filtered Graph\footnote{A Planar Maximally Filtered Graph is a filtered graph obtained by subsequently adding edges to a graph initially deprived of all its edges. Edges are added in a descending order accordingly with their weight, if and only if the resulting graph can still be embedded on a planar surface. A Planar Maximally Filtered Graph preserves the hierarchical organization of the Minimum Spanning Tree but contains a larger amount of information in its structure and proves to be efficient in filtering relevant information about the clustering of the system.} (PMFG) proposed in \cite{Tumminello10421}. We then define different topics  {as} clusters of mathematical concepts identified by the Louvain community detection algorithm \cite{Louvain} over the PMFG graph. Table \ref{tab:2} shows the number of topics extracted for each project and the modularity of the partition of keywords in the filtered co-occurrences network. Using such definition of topics, we label each post with the topic whose associated keywords are most present in the text. In case of a tie, no label is associated to the post. Therefore, in addition to its time of publication ($t$), its author ($\alpha$) and its content ($K$), each post is also characterized by a topic label ($T$): $\Pi_i=(t_i,\alpha_i,K_i, T_i)$.
\begin{table}[h!]
\centering
\label{tab:2}       
\begin{tabular}{lll}
\hline\noalign{\smallskip}
Project & Topics & Modularity \\
\noalign{\smallskip}\hline\noalign{\smallskip}
1 & 9 & 0.70 \\
4 & 9 & 0.75\\
5 & 10 & 0.72\\
8 & 12 & 0.72\\
15 & 9 & 0.69 \\
\noalign{\smallskip}\hline
\end{tabular}
\caption{Number of topics per project}
\end{table}

\subsection{Similarity and Innovation}
\label{innovDef}
We first define the semantic similarity between two posts using the Jaccard measure between their contents: $J_{ij}=(K_i\cap K_j)/(K_i\cup K_j)$.\\
 We tweak this similarity by taking into consideration the temporal distance among the posts, thus introducing the \emph{semantic-temporal} similarity:
\begin{equation}
    \Theta_{i,j}=J_{i,j}e^{-|t_i-t_j|/\tau_0}
\end{equation}
where $\tau_0$ is a the average time distance among all the couples of posts, that is therefore different for each project.  According to this measure, two posts similar in content but distant in time, will be less similar than according to the standard Jaccard measure.\\
We use the semantic-temporal similarity measure to define an innovation index for each post. First we define for each post two separate indicators: 
\begin{itemize}
    \item The \emph{in-debate index} measures how a post is similar to the contents published just before it. It is calculated as the average of the semantic-temporal similarity from the previous posts:
    \begin{equation}
        \nu_i=\frac{\sum_{j|t_j<t_i}\Theta_{ij}}{\sum_{j|t_j<t_i}1}
    \end{equation}
    \item The \emph{impact index} measures how a post content is reproduced in the posts following it. It is calculated as average Jaccard similarity with the following posts:
    \begin{equation}
        \xi_i=\frac{\sum_{j|t_j>t_i}J_{ij}}{\sum_{j|t_j>t_i}1}
    \end{equation}
\end{itemize}
An innovative post is characterized by a low in-debate index (i.e. it is different from with the earlier content) and a high impact (it influences the following contents that are therefore similar with it). For this reason we define the \emph{innovation index}, for each post:
\begin{equation}
    I_i=-\xi_i log(\nu_i).
\end{equation}
\section{Results}
\label{sec:2}
\subsection{Organization of labor}
\label{laborOrg}
\begin{figure*}
\includegraphics[width=1\textwidth]{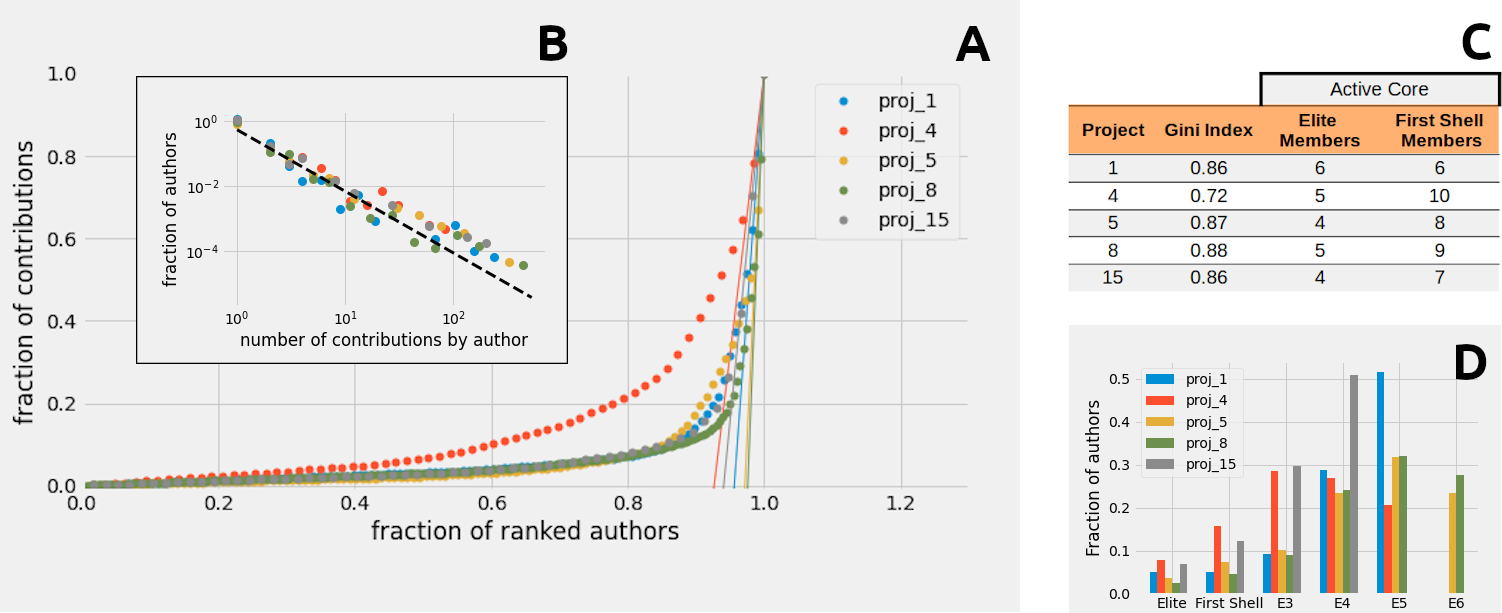}
\caption{A: Lorenz curve for authors' contributions to the projects. B: Empirical probability distribution of the number of contributions by author. C: Statistics about contributors activity. D: Elite--peripheries structure of the projects.}
\label{heterogeneity}       
\end{figure*}

As usually happens in collaborative systems, few contributors do most of the work \cite{barabasi}. Analyzing the number of posts produced by each author, we notice a power--law distribution of the number of contributions by author (Fig.~\ref{heterogeneity}B) and elevates Gini indices (Fig.~\ref{heterogeneity}C). In Fig.~\ref{heterogeneity}A we represented this heterogeneity under the form of the Lorenz curve: authors are ranked according to the number of contributions and curves represent the cumulative fraction of contributions produced by the corresponding fraction of ranked authors.
We can observe from the figure that the most active 10\% produces the 80\% of the posts (except for project 4, also characterized by a lower Gini index, where the 20\% of the contributors produces the 80\% of the posts). \\
Following the procedure in \cite{bassolas}, we use the Lorenz curve to hierarchically classify groups of authors: we take the derivative of the Lorenz curve at  the point (1,1) and we set a first threshold at the point where the derivative crosses the x axis (as you can see in Fig. \ref{heterogeneity}A). The authors after this threshold represent the most productive \emph{elite} of the project. We remove this \emph{elite} contributors and we recursively repeat the procedure, identifying, at the first iteration, a group that we define as the \emph{first shell} (highly active contributors but outside the hyper-active \emph{elite}) and at the following iterations the \emph{peripheral shells} (namely E3, E4, E5, E6 shells in Fig. \ref{heterogeneity}D). In Fig.~\ref{heterogeneity}C we display the number of contributors in the elite group and in the first shell, while Fig.~\ref{heterogeneity}D shows the percentage of authors in each hierarchical category. We can see that, according to this classification, the elite group comprises less than 10\% of the authors while the peripheral shells are consistently the most represented. In the following we will refer to authors belonging to the \textit{elite} and the \textit{first shell} as the \textit{active core}. 

\subsubsection{Authors' interactions}
\begin{figure*}[h!]
\includegraphics[width=\textwidth]{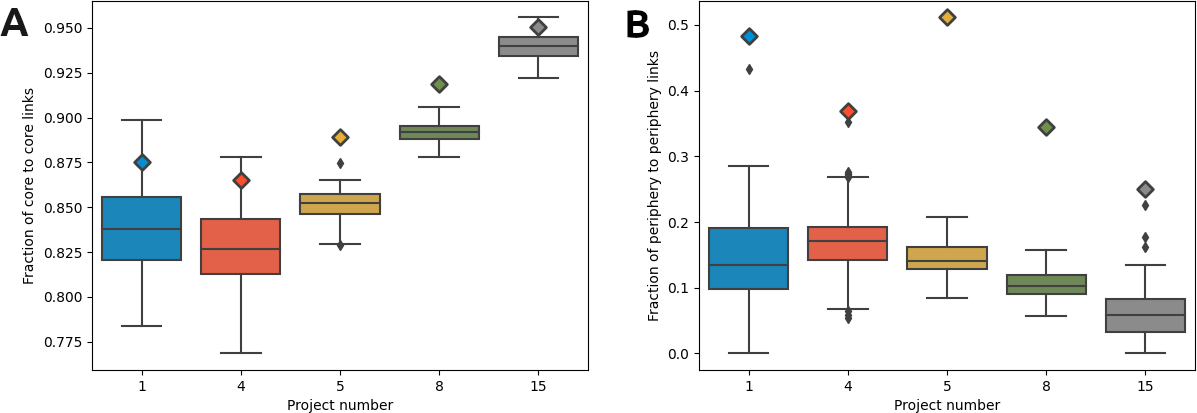}
\caption{\textbf{Interactions though comments:} A: Boxplots represent the distribution of the fraction of in-core links after 100 simulations. $ \bdiamond$ represents the fraction of in-core links in the Polymath dataset. B: Boxplots represent the distribution of the fraction of in-periphery links after 100 simulations. $ \bdiamond$ represents the fraction of in-periphery links in the Polymath dataset.}
\label{fig:comments}       
\end{figure*}

To better understand the division of labor in Polymath, we investigated the distribution of interactions between authors. In particular, we focused on how the \textit{active core} authors, as defined in the previous section, interact with the \textit{peripheral shells}. 
\newline
In order to do so, based on the dependencies between posts, we defined a \textit{comments interaction network} CIN $= (\mathcal{V}, \mathcal{E}, W)$ with the following properties: each node $i \in \mathcal{V}$ represents an author, an edge $(i,j) \in \mathcal{E}$ represents the existence of at least one comment by author $i$ to a post of author $j$, and the weight $W_{ij}$ associated to the edge $(i,j)$ represents the number of times author $i$ replied to a post of author $j$.
To understand weather such interactions are highly concentrated in the active core of the elite contributors or more spread towards peripheral contributors we compared the obtained graphs with a stochastic network formation model preserving, on average, the level of activity of each node. Similarly to \cite{ROTH2013394}, we hence simulate $K$ networks $\{ \Gamma^k = (\mathcal{V}, \mathcal{E}^k, Q^k)\}_{k \in \{1,\ldots,K\}}$ with an expected degree for each node equal to the one of the authors of our dataset, keeping the same number of nodes $n = |\mathcal{V}|$ and edges $m = |\mathcal{E}| = |\mathcal{E}^k| \:$ for all $k \in \{1,\ldots,K\}$. To do so we draw the weights $Q_{ij}^k$ from a multinomial distribution with parameters $m$ and $p = \{p_{ij}\}_{i,j \in \mathcal{V}}$ such that
\[
p_{ij} = \frac{d_i^{\text{out}}\cdot d_j^{\text{in}}}{m^2}
\]
where $d_i^{\text{out}}$ is the out-degree of node $i$ and $d_j^{\text{in}}$ is the in-degree of node $j$ in the comments interaction network.
\newline
Fig. \ref{fig:comments}A shows the distribution of the \textit{fraction of in-core links} (namely the fraction of messages by elite contributors directed to other elite contributors) in our $K = 100$ simulations and compares such distributions with the actual fraction of in-core links in our dataset. Fig. \ref{fig:comments}B, proposes the same comparison for the in-periphery links, namely the fraction of messages written by peripheral contributors directed to other peripheral contributors.  
Both plots reveal a peculiar division of labor in the Polymath project: both core-to-core and periphery-to-periphery links are more represented than they are in random simulations, underlining that authors are more inclined to reply to contributors who participate in the discovery process to the same extent.
\newline

\begin{figure*}[h!]
\includegraphics[width=1\textwidth]{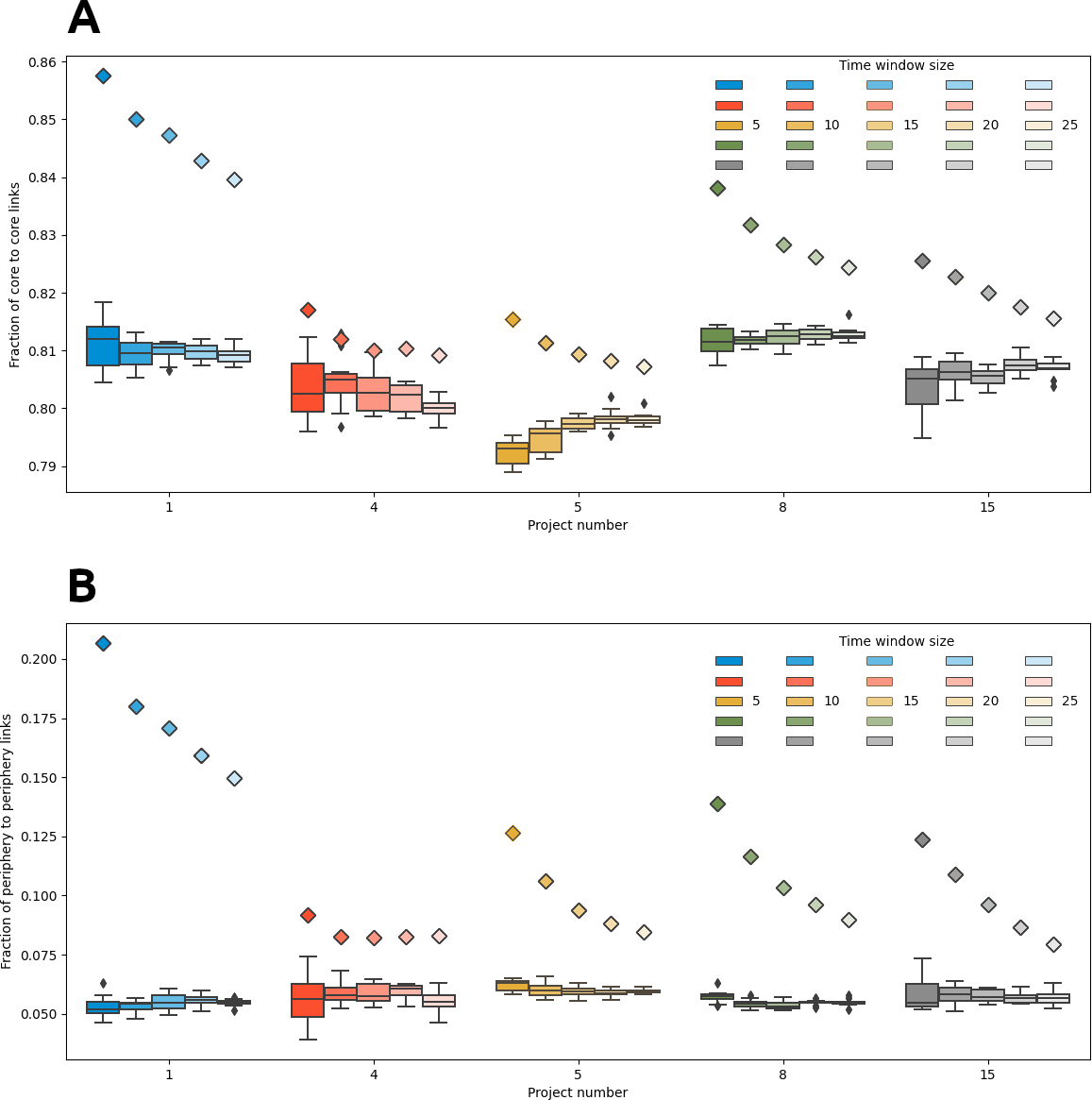}
\caption{\textbf{Interactions on the same topic:} boxplots represent the outcome of simulations while $ \bdiamond$ markers represents real values coming from data. A: Percentage of links stemming from a core node and ending in a core node over the total number of edges stemming from core nodes. B: Percentage of links stemming from peripheral nodes and ending in peripheral nodes over the total number of edges stemming from peripheral nodes. The increasing transparency corresponds to the increase of the time window.}
\label{fig:topics}       
\end{figure*}

As pointed out in the Data and Methods section though, some of the studied blogs presents constraints to the comments network development. Moreover, we qualitatively observed a shift from non hierarchically structure in the very first project (i.e. solely presence of second layer comments and no deeper structures) to a more articulated organizations of the posts in the latest projects. Hence, to evaluate the robustness of the results presented in Fig. \ref{fig:comments}, we compared them to the ones obtained through a different definition of network interactions. We define a \textit{topic interaction network} TIN(T) $= (\mathcal{V}, \widetilde{\mathcal{E}}(T), \widetilde{W}(T))$ with the following properties: the node set $\mathcal{V}$ still represents the set of authors, an edge $(i,j) \in \widetilde{\mathcal{E}}$ represents the fact that authors $i$ and $j$ published a post on the same topic at a distance no bigger than $T$ posts (when posts are ordered chronologically). The weight $\widetilde{W}_{ij}(T)$ associated to the edge $(i,j)$ represents the number of times author $i$ and $j$ published a post on the same topic in the time window defined by parameter $T$. Notice that, by definition, such a network is undirected. 
Once again, in order to study authors interactions, we need to compared them with a set of simulated networks $\{ \widetilde{\Gamma}^k = (\mathcal{V}, \widetilde{\mathcal{E}}^k, \widetilde{Q}^k)\}_{k \in \{1,\ldots,K\}}$ where $\widetilde{m} = |\widetilde{\mathcal{E}}| = |\widetilde{\mathcal{E}}^k|,\; k \in \{1, \ldots, K\}$. To do so, it is now sufficient to draw the solely values $\{\widetilde{Q}_{ij}\}_{i,j \in \mathcal{V}, j \geq i}$ from a multinomial random distribution, as we want the network to be undirected and $\widetilde{Q}_{ij}^k = \widetilde{Q}_{ji}^k$ for all the $K$ simulations. Therefore, we draw the values $\{\widetilde{Q}_{ij}\}_{i,j \in \mathcal{V}, j \geq i}$ from a multinomial distribution of parameters $\widetilde{m}$ and $\widetilde{p} = \{\widetilde{p}_{ij}\}_{i,j \in \mathcal{V}, j \geq i}$ such that
\begin{align*}
\widetilde{p}_{ij} &= 2\frac{d_i\cdot d_j}{\widetilde{m}^2} \;\;\;\;\;\;\;\;\;\;\;\; \text{if } i \neq j \\
\widetilde{p}_{ii} &= \frac{d_i\cdot d_i}{\widetilde{m}^2} \;\;\;\;\;\;\;\;\;\;\;\;\;\; \text{otherwise,}
\end{align*}
where $d_i$ is the degree of node $i$ in the topic interaction network.
The resulting distribution of in-core and in-periphery interactions is shown by Fig. \ref{fig:topics}. 
We can notice that, no matter how we chose the tie window, in-core and  in-periphery interactions are higher in the actual Polymath projects than in simulated networks. Moreover, results obtained on the \textit{topic interaction network} confirm the ones obtained on the \textit{comment interaction network}. Thus, we can solidly conclude that
in the Polymath collaborations, elite actors interact more with other elite actors, while peripheral actors, only sporadically contributing to the development process, respond to other peripheral actors more than in a random model preserving their level of activity. This result is in agreement with a characteristic often  observed in social network structures, that is status-based homophily \cite{mcPherson}. It is however surprising in the scientific context where interactions are usually assumed to be based on cumulative advantage processes \cite{merton}.

\subsection{Collective intelligence at work}
\label{collInt}
\begin{figure*}
\includegraphics[width=1\textwidth]{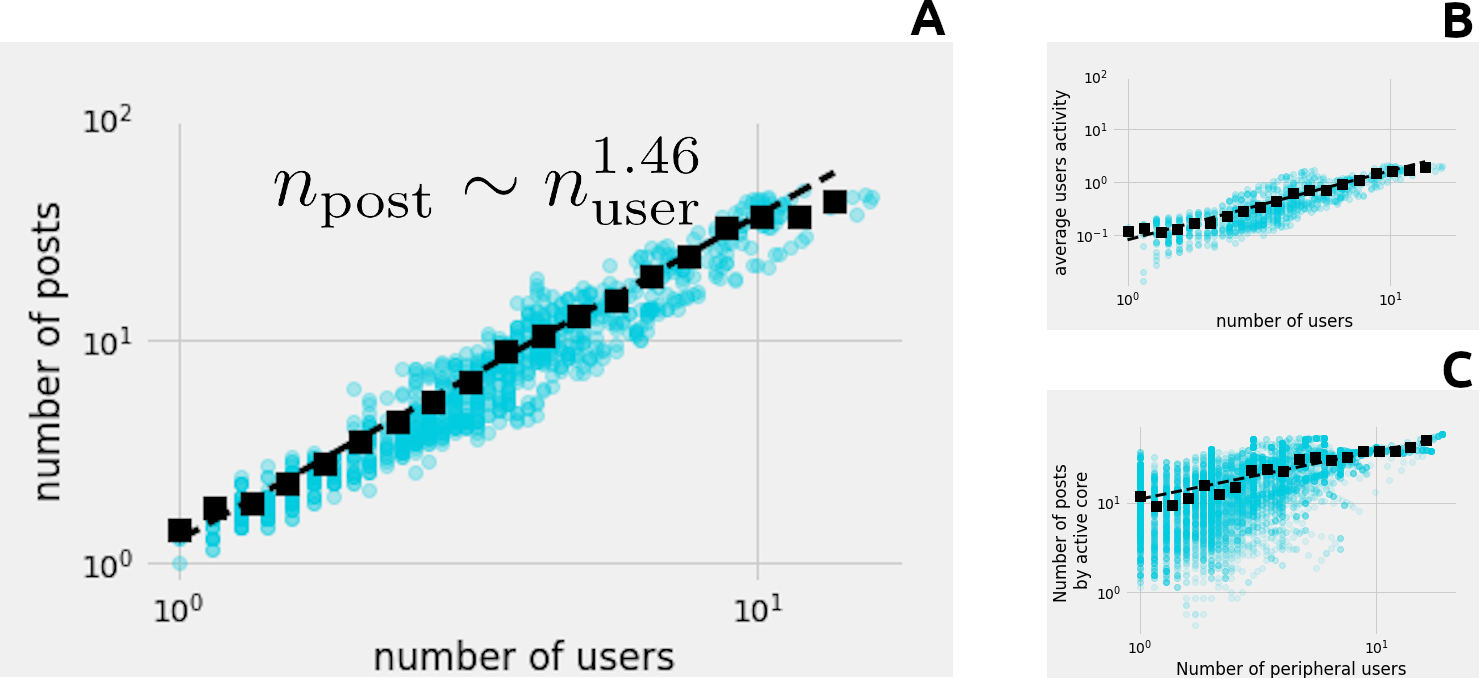}
\caption{A: Superlinear growth of the number of posts with the number of users. B: Average daily individual productivity as a function of the number of users. C: Average daily individual productivity of core authors as a function of the number of peripheral contributors.}
\label{super}       
\end{figure*}

Several studies on collaborative systems showed the superlinear effect of collaboration: collective intelligence means that when more people are involved in the project the total productivity (in our case the number of posts) is higher than the sum of the individual productions. To test this feature dynamically, for all the projects, we count the daily number of posts and the daily number of participants:
\begin{equation}
    n_{post}(t)=[n_{post}(t_0),n_{post}(t_1),...];\quad n_{user}(t)=[n_{user}(t_0),n_{user}(t_1),...]
\end{equation}
where $t_0,t_1,\ldots$ represent different days.
To reduce the noise, we smooth these time series using a 7-days rolling window. 
Plotting the couples $(n_{user}(t),n_{post}(t))$ for all the days we obtain the curves representing the relationship between the number of users and the number of posts. In Fig.~\ref{super}A we show the evident superlinear growth, $n_{post}=n_{user}^\gamma$ , (with exponent $\gamma=1,46$) of the number of posts with the number of users, aggregated on all the projects. Our results are similar to those obtained in \cite{sornette} for GitHub. \\
Fig.~\ref{super}B and Fig.~\ref{super}C suggest that contributions have positive superlinear effects, even when they are relatively marginal. In Fig.~\ref{super}B we show that the average individual daily production (for all the contributors with more than 10 posts in all the project) grows with the number of users active in that day. Fig.~\ref{super}C displays the daily average productivity of the active core as a function of the number of users in the peripheral shells. We observe that an important presence of peripheral users boosts the productivity of the most active users. \\
 In Fig.~\ref{super} we showed the results obtained aggregating together all the projects. The individual analysis of each blog shows similar trends with very small variations of the growth exponents (blog1: $\gamma=1.30$, blog4: $\gamma=1.22$, blog5: $\gamma=1.46$, blog8: $\gamma=1.65$, blog15: $\gamma=1.50$). Being the blog platforms quite different among them, for what concerns the users' interaction structure (comments' dependencies), the robustness of this results among the blogs also means that we can consider super--productivity an intrinsic characteristic of collaborative science, independently by the communication medium.

\subsection{The statistical properties of scientific discoveries}
\label{statProp}
\begin{figure*}
\includegraphics[width=1\textwidth]{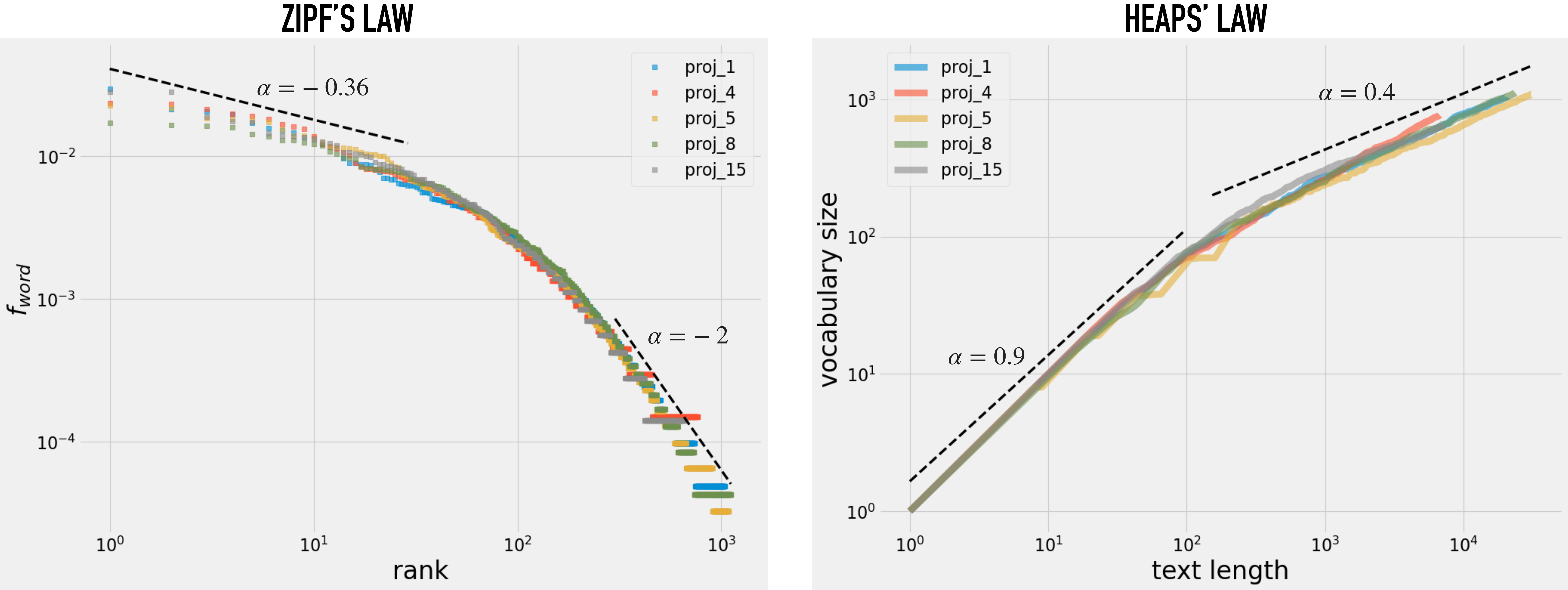}
\caption{Left plot: the rank-frequency relationship respects the Zipf's law with the exponents indicated in the plot. Right plot: the relationship between the text length and the vocabulary size respect the Heaps' law with the exponents indicated in the plot.}
\label{linguistic}       
\end{figure*}

While in the previous section we tried to understand the collaborative patterns of open science, we now focus on the analysis of the scientific discovery process itself. \\
We first analyze the statistical properties of the mathematical concepts  mobilised in the projects. As described in the methods section, we  associated each post to a set of mathematical concepts. We first test if our corpus follows the basic laws of linguistic patterns: Zipf's law and Heaps' law. Zipf’s law expresses the relationship between the frequency and the ranking of words. It states that the frequency of a word is inversely correlated with its rank, $f\sim r^{-\alpha}$. For example, when considering the Gutenberg text corpus, which is a large sample of the English literature production, it has been observed a value $\alpha\sim -1$ for low values of $r$ and $\alpha\sim -2$ for high values of $r$. Heaps’ law concerns the entrance of innovative concepts in a text and it express the relationship between the number of different words (vocabulary size) and the total number of used words (length of the text). It describes an initial linear growth followed by power law asymptotic behavior, $l=v^\alpha$: in the Gutenberg corpus it has been observed $\alpha\sim 1$ for low values of $l$ and $\alpha\sim 0.44$ for high values of $l$. 
In Fig. \ref{linguistic}, we observe that not only these laws are respected in our corpora, but also all the project show the same behavior and display the same exponents. The first exponent of Zipf's law is lower compared with the Gutenberg corpus due to the fact that we removed the non-mathematical common expressions (notably the stop words). This signifies that the creative process of the scientific discovery follows the same basic rules of literary production. 

\begin{figure*}
\centering
\includegraphics[width=0.6\textwidth]{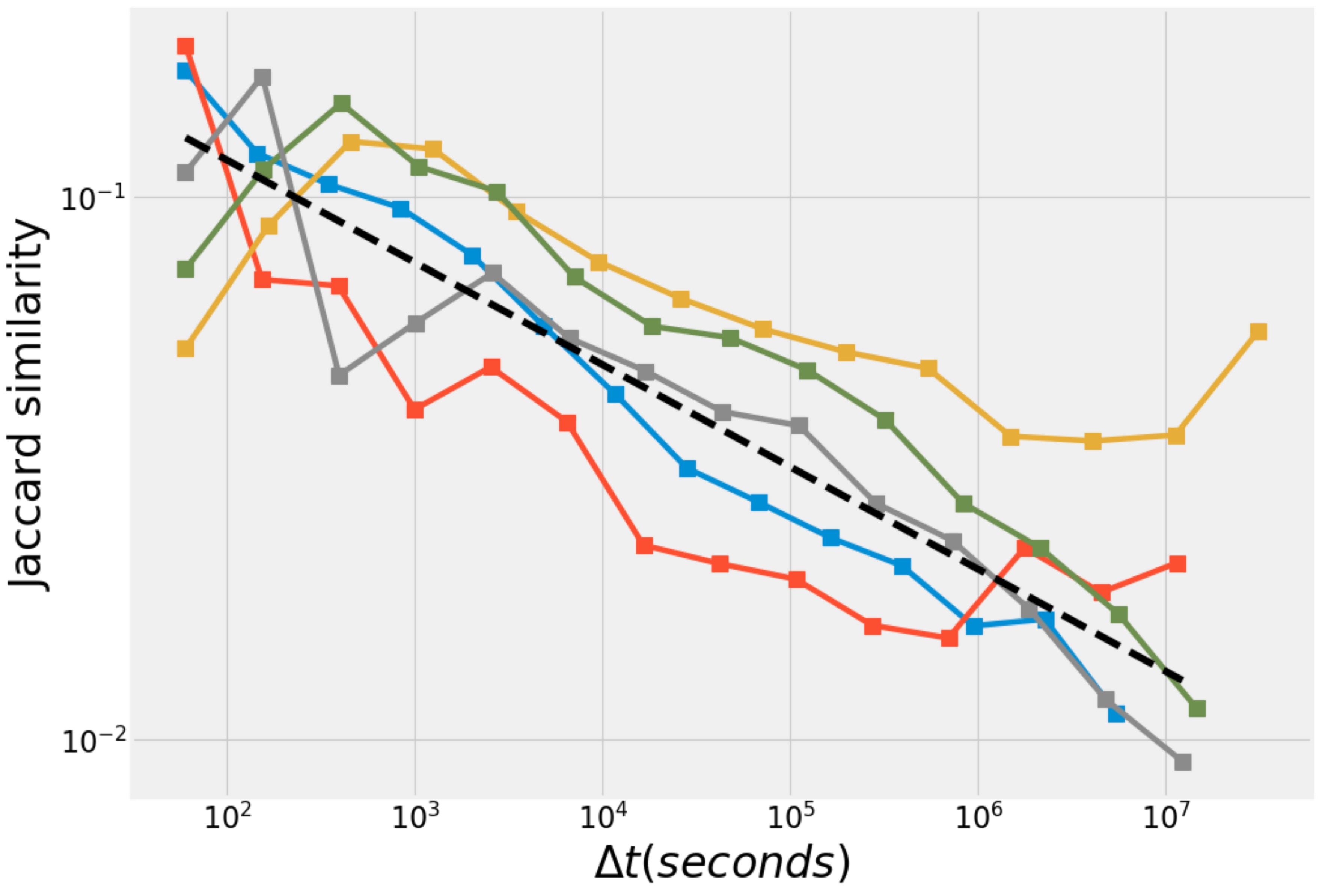}
\caption{Average distance between posts at a certain temporal distance.}
\label{jaccardTime}       
\end{figure*}

Secondly, we focused on the typical time of the discovery process, arguing that the closer two posts are in time the more similar they should be in terms of content. In Fig. \ref{jaccardTime} we display the average Jaccard similarity between all the couple of posts published with a certain time lag. We observe indeed a power law decay of the similarity with time, $J\sim\Delta t^{-\gamma}$ (with $\gamma=0.2$), once again  similar for all the projects. Hence, in all projects, there exists a typical time window for which the debate remains focused on the same topic before changing to another one. 

\subsection{Innovation patterns}
\label{innovPatterns}
\begin{figure*}
\includegraphics[width=1\textwidth]{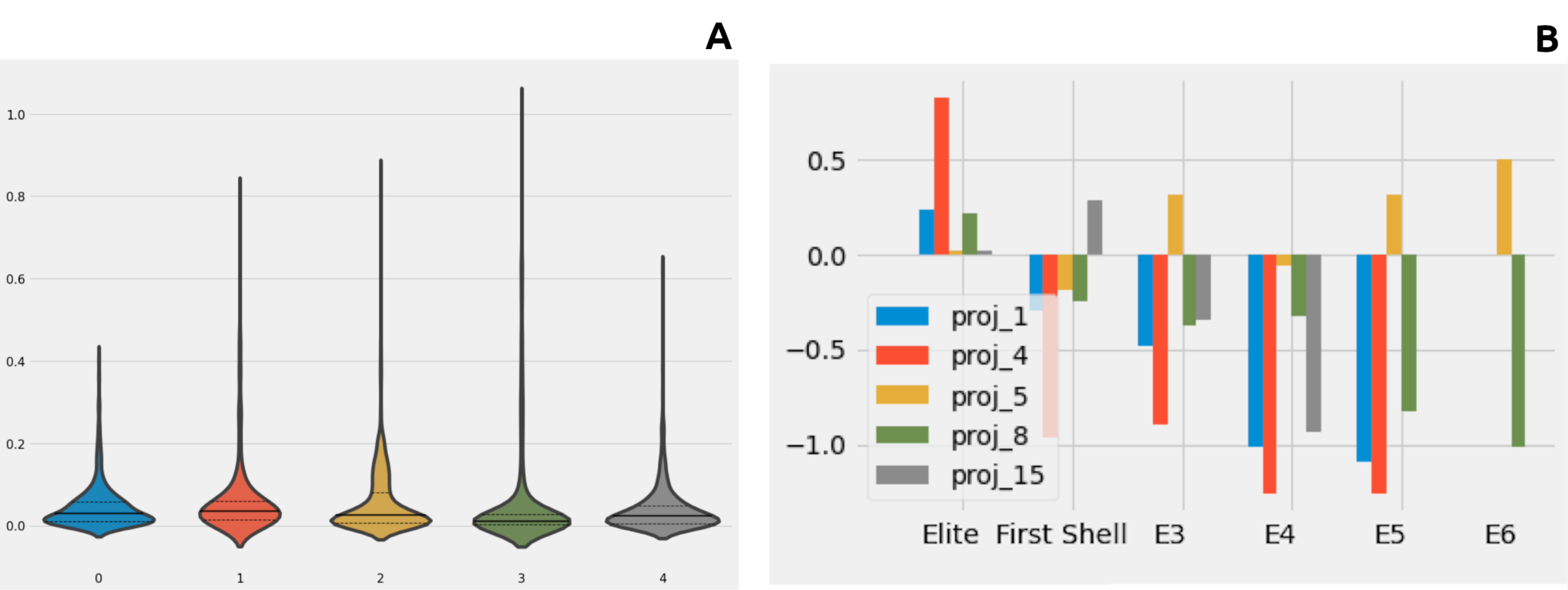}
\caption{Plot A: Distributions of innovation for each project.Plot B: Innovation z-score for each activity shell. }
\label{innovation}       
\end{figure*}

Finally we analyze how innovations affect the discovery mechanism, using the innovation measure we defined in Section~\ref{innovDef}. As we can observe in Fig.~\ref{innovation}A the innovation values' distributions are long tailed, meaning that few posts have a much larger innovative content compared to the others: high innovation is rare, but statistically significant. \\
Defining \textit{innovative} those posts that are in the upper quartile of the innovation distribution and recalling the definition of the activity shells that we introduced in Section \ref{laborOrg}, we now analyze which actors  are the main drivers of innovation. Since groups' sizes are different, we compare the number of innovations observed in each class with its multinomial expectation, namely the probability of a post to be innovative (25\%), multiplied by the number of posts produced by the group. We calculate the z-score between the observed and the expected values. While the previous results showed a quite homogeneous behavior between the different projects, here we observe marked differences. 
In projects 1,4,8 the elite produces more innovation than expected. The first shell is the most important innovation driver in project 15.  Finally, in projects 5, the peripheral shells are the largest producers of innovation. This result puts in evidence that, in large scale collaborations, there is no strict rule determining, a priori, who the major innovators will be and sometimes serendipitous interactions by peripheral contributors can have a large impact on the discovery process: a single post by a occasional participant can be responsible of opening a large adjacent possible and giving a new direction to the work. 

\section{Discussion}

In the last decades we have assisted to the rise of major online collaborations, among which it's worth recalling Linux, GitHub  and Wikipedia. In 2009 Polymath projects took off with the specific purpose of exploiting online collaborative environments to solve mathematical issues, letting scientific research embrace massive online collaboration for the first time.  \\
In this work we investigated how the path to scientific discovery develops in such collaborative environments, how the labor is organized between authors and which actors play the role of innovators.
Our results, consistently with previous work, show that productivity is highly skewed between contributors and identify a small hyper-productive elite publishing the largest majority of the posts.
Nevertheless, peripheral contributors do have a crucial role in fostering the overall production: in fact, content production grows super-linearly with the number of participants in the discussion.  Interestingly, our analysis shows that, in Polymath projects, peripheral contributions boost other authors activity in a rather indirect way.  Although interactions between the elite and the rest of the participants are relatively limited, as both peripheral and hyper-productive authors are inclined to reply to authors with similar level of activity, we proved that peripheral contributors often have an important effect in  bringing innovation and new ideas into the debate. In fact, we pointed out that there is no general rule determining who the main innovators are based on their level of production. Sometimes sporadic contributors can play a significant role in innovating the discussion and can be credited with leading the research to new directions.  

It would be interesting, in future works, to compare the results obtained on the Polymath dataset with the ones regarding some other online collaborative environment, especially analyzing the relation between level of participation and innovation, as not much is available in literature so far.


\section*{Funding}
This research has been supported by CNRS through the 80 PRIME MITI project ``Disorders of Online Media'' (DOOM).



\bibliographystyle{bmc-mathphys} 
\bibliography{bmc_article}      



\end{document}